%% file: iclr2027_conference_preprint.tex
\documentclass{article} 
\usepackage{iclr2027_conference,times}

\input{math_commands.tex}

\usepackage{hyperref}
\usepackage{url}
\usepackage{booktabs}
\usepackage{multirow}
\usepackage{graphicx}
\usepackage{wrapfig}
\usepackage{placeins}   
\usepackage{xcolor}
\usepackage{amsmath,amssymb}
\usepackage{tikz}
\usetikzlibrary{arrows.meta,positioning,fit,backgrounds,calc,shapes.geometric}
\usepackage{pgfplots}
\pgfplotsset{compat=1.17}
\usepackage[most]{tcolorbox}
\usepackage{enumitem}
\definecolor{teal}{HTML}{0F6B67}
\definecolor{violet}{HTML}{6A4BB0}
\definecolor{burnt}{HTML}{B4541A}
\definecolor{soft}{HTML}{475553}

\newcommand{\ours}{\textsc{Listen-to-Reason}}
\newcommand{\oursacc}{\textsc{L2R}}

\title{LISTEN-to-Reason: Listen with Experts, Retrieve over a Graph, Reason with LLMs}

\author{Pooneh Mousavi$^{1,2}$\thanks{Corresponding author: Pooneh Mousavi.}, \ Mirco Ravanelli$^{1,2}$, \ Cem Subakan$^{1,2,3}$ \\
\normalfont $^{1}$Concordia University \quad $^{2}$Mila -- Quebec AI Institute \quad $^{3}$Laval University}

\iclrfinalcopy 
\newcommand{\projecturl}{https://poonehmousavi.github.io/listen-to-reason/}

\newcommand{\projectexamples}{\url{https://poonehmousavi.github.io/listen-to-reason/}}
\begin{document}

\maketitle
\lhead{Preprint}
\input{sections/abstract}
\input{sections/intro}

\input{sections/related}
\input{sections/method}

\input{sections/setup}
\input{sections/results}
\input{sections/conclusion}

\input{sections/statements}

\bibliography{iclr2027_conference}
\bibliographystyle{iclr2027_conference}

\appendix
\input{sections/appendix}

\end{document}

%% file: math_commands.tex
\usepackage{amsmath,amsfonts,bm}

\def\eqref#1{equation~\ref{#1}}

\def\1{\bm{1}}

\DeclareMathAlphabet{\mathsfit}{\encodingdefault}{\sfdefault}{m}{sl}
\SetMathAlphabet{\mathsfit}{bold}{\encodingdefault}{\sfdefault}{bx}{n}



%% file: sections/abstract.tex
\begin{abstract}

Large audio-language models (LALMs) fuse an audio encoder into a large language model (LLM) through multi-stage training. This coupling means that a new domain or a stronger LLM requires retraining, and their answers cannot be traced to what the model heard: a chain-of-thought is a post-hoc account. We propose Listen-to-Reason (L2R), an interpretable-by-design pipeline that passes audio to the LLM through an explicit, human-readable tree: small heads on frozen expert encoders map each chunk of a clip to semantically meaningful nodes on the tree (for speech, music and environmental sound), and a frozen text-only LLM answers from these nodes and an ASR transcript without hearing the clip. Every answer can therefore be traced to the nodes and transcript it read, and the nodes are causal: replacing the deciding node with a distractor overturns 78\% of correct answers on SAKURA. With a 7B reader, L2R outperforms all LALMs we compare against on SAKURA and trails them by 6--12 points on MMAU and MMAR, despite training about 1{,}400$\times$ fewer parameters on orders of magnitude less audio data. However, because any LLM can serve as the reader, we show that a stronger reader narrows this gap without retraining any audio component.
A new domain is added with one small head: with five labelled clips per species, it outperforms QLoRA fine-tuning of an LALM on the same clips by 13--26 points. Examples and an interactive view of the tree are available on our \href{\projecturl}{project page}.

\end{abstract}

%% file: sections/intro.tex
\section{Introduction}
\suppressfloats[t]
\label{sec:intro}

\input{sections/fig_pipeline}

Large audio-language models (LALMs) extend large language models (LLMs) to audio \citep{cui2025recent,arora2025landscape}.
A typical LALM encodes the audio with a pretrained encoder, maps the encoder output into the input space of the LLM through an adaptor, and trains the combined model on audio-text data \citep{deshmukh2023pengi,gong2024ltu,tang2024salmonn,chu2023qwenaudio,chu2024qwen2audio,ghosh2024gama,kong2024audioflamingo}.
They range from models that answer in text \citep{gong2023ltuas,hu2024wavllm,ghosh2025af2,goel2025af3} to models that also speak \citep{zhang2023speechgpt,defossez2024moshi,zeng2024glm4voice,xie2024miniomni,fang2025llamaomni,wang2025freezeomni,yu2025salmonnomni} and omni models that add vision \citep{fu2024vita,abouelenin2025phi4mini,kimiteam2025kimiaudio,huang2025stepaudio,xu2025qwen25omni,xu2025qwen3omni}.

These abilities come from long, multi-stage training: the audio encoder is first aligned with the LLM on perception data such as automatic speech recognition (ASR) and audio captioning, then tuned on instruction and question-answer pairs, and finally taught to reason with chain-of-thought (CoT) data or reinforcement learning (RL) \citep{tang2024salmonn,gong2024ltu,goel2025af3,xie2025audioreasoner,xu2025qwen3omni}.
Each stage updates the adaptor, and often the encoder and the LLM as well, with large data and compute budgets \citep{gong2024ltu,goel2025af3,xu2025qwen25omni,xu2025qwen3omni}.
The encoder and the LLM are therefore coupled: replacing either one, with a stronger LLM or a better encoder, means repeating the training.
Adapting the model to a new domain likewise requires another fine-tuning run, which can degrade what it already knows \citep{luo2023forgetting}.

Recent benchmarks test audio reasoning directly: MMAU \citep{sakshi2024mmau}, MMAU-Pro \citep{kumar2025mmaupro}, MMAR \citep{ma2025mmar} and SAKURA \citep{yang2025sakura} ask questions over speech, sound and music that need expert knowledge or multi-step inference.
To do well on them, LALMs are trained to write a CoT \citep{wei2022cot} before answering, by supervised fine-tuning on reasoning traces or by RL \citep{xie2025audioreasoner,kong2025soundcot,li2025r1aqa,wen2025sari,rouditchenko2025omnir1,wu2026audiothinker,wijngaard2025audsemthinker,tian2025stepaudior1,xiong2025thinking}, and models with such thinking modes, including Audio Flamingo 3 and Qwen3-Omni, are among the strongest open models \citep{goel2025af3,xu2025qwen3omni}.
CoT does not always help, however: it can lower accuracy on hard items and shift attention away from the audio \citep{ma2025audiocot,bonzi2026spare}.

Whether CoT helps or not, it does not explain the answer.
In text LLMs, CoT explanations can be unfaithful to the decision they accompany \citep{turpin2023unfaithful,lanham2023faithfulness}.
In LALMs, reasoning chains often agree with the prediction but are weakly grounded in the audio, and they change under hallucination-inducing and adversarial inputs \citep{mousavi2026faithfulness}.
The trace is generated by the same network that makes the decision, after the audio has been fused into its hidden states, so it is a post-hoc account that can change with the prompt or the sampling.
Learning what an LALM actually used requires separate analyses, such as alignment scores, layer-wise probes, cue-reliance tests and hallucination benchmarks \citep{mousavi2025alas,xiang2025modality,yang2025audiolens,chen2026audio,foo2026glitters,kuan2024understanding,kuan2025truly}; these sit outside the model, whose own output does not state what it heard.

A different line of work puts the evidence in the model's input.
Tool-using LLMs call other models, programs or APIs and read their outputs \citep{schick2023toolformer,yao2023react,shen2023hugginggpt,lu2023chameleon,patil2024gorilla,qin2024toolllm,wu2023visualchatgpt,yang2023mmreact,gupta2023visprog,suris2023vipergpt}, and audio agents apply this idea to audio understanding \citep{huang2024audiogpt,audiotoolagent2025,wang2026audiomind}.
Symbolic cascades describe the audio in text and let a text LLM reason over the description \citep{zeng2022socratic}; closest to our work, SAR-LM converts audio into human-readable features across speech, sound and music and uses them to trace errors, but it concatenates the separate label spaces of off-the-shelf extractors rather than one vocabulary \citep{taheri2025sarlm}.
Retrieval-augmented generation (RAG) conditions the answer on retrieved documents \citep{lewis2020rag,gao2023ragsurvey}, and graph-based RAG retrieves and traverses a graph built over the knowledge source \citep{edge2024graphrag,gutierrez2024hipporag,he2024gretriever,peng2025graphrag}.
Retrieval improves accuracy on questions that need outside knowledge, and it improves interpretability, because the evidence the model conditions on is part of its input and can be read.

Knowledge graphs (KGs) give this evidence an explicit structure.
They store facts as typed nodes and edges, and combining them with LLMs grounds generation in these facts and exposes the reasoning path \citep{pan2024unifying,jiang2023structgpt,baek2023kaping,sun2024thinkongraph,luo2024rog}.
In vision, multimodal KGs, scene graphs and retrieved knowledge support visual question answering and multimodal reasoning \citep{liu2019mmkg,wang2020richpedia,alberts2021visualsem,zhu2022mmkgsurvey,krishna2017visualgenome,marino2019okvqa,marino2021krisp,gui2022kat,chen2022murag,hu2023reveal,lee2024mrmkg,abootorabi2025askany}.
Audio has few counterparts.
iKnow-audio refines the zero-shot predictions of CLAP, a contrastive audio--text model rather than an LLM, with a knowledge graph of relations between sound events \citep{olvera2025iknowaudio}, and Audiopedia links the transcript of a clip to Wikidata entities and adds their facts to the prompt of an LALM \citep{penamakuri2025audiopedia}.
Graph retrieval has used audio only as one track of video \citep{park2025m3kg}, and audio RAG retrieves clips by embedding similarity \citep{chen2025wavrag}.
To our knowledge, no multimodal KG covers speech, music and environmental sound with nodes that a clip can be routed to directly from its waveform.

Such a graph would make audio understanding interpretable by design.
A model is interpretable when a person can see why it made a decision \citep{doshivelez2017rigorous,lipton2018mythos}.
Post-hoc methods explain a trained black box, such as an audio classifier \citep{paissan2024lmac,paissan2024lmaczs}, after the fact, and their explanations can misrepresent it: saliency maps can be insensitive to the model's weights, and attention weights are not reliable explanations \citep{adebayo2018sanity,jain2019attention,wiegreffe2019attention,rudin2019stop}.
A model that is interpretable by design decides from an intermediate representation a person can read, so the explanation is the computation itself \citep{rudin2019stop,koh2020cbm,chen2019thislooks,oikarinen2023lfcbm}.
This matters where errors are costly \citep{topol2019highperformance}, as in healthcare: clinicians need to check the evidence behind a prediction \citep{tonekaboni2019clinicians,amann2020health,ghassemi2021falsehope}, speech is increasingly used to assess depression, Parkinson's and Alzheimer's disease \citep{cummins2015review,tsanas2012parkinson,luz2020adress,low2020psychiatric,fagherazzi2021voice,botelho2024speechbiomarker}, and audio models used there should state the evidence behind each decision \citep{akman2024audioxai,chen2026voicecbm}.

We present \ours{} (\oursacc), {which bottlenecks the audio representation through} a multimodal knowledge-tree. This tree connects audio to an LLM and makes the evidence behind each answer explicit (Figure~\ref{fig:pipeline}).
The tree describes a clip in named nodes that are grounded in audio, and a frozen text-only LLM answers from these nodes, together with an ASR transcript of any speech or singing, without hearing the clip.
The encoders and the LLM communicate only through text, which decouples perception from reasoning: each can be replaced or extended without retraining the other.
Our contributions are as follows:
\begin{itemize}[leftmargin=*,itemsep=2pt]
  \item \textbf{Introducing \ours{}.} We propose \oursacc{}, an audio question-answering model in which the audio reaches the LLM only as text: named tree nodes (\S\ref{sec:graph}) and an ASR transcript. A router of small heads on frozen expert encoders traverses the tree for each chunk of a clip and retrieves the nodes it hears as a readable context (\S\ref{sec:router}), and a frozen text-only LLM answers from these nodes and the transcript without hearing the clip (\S\ref{sec:reader}). Any LLM can serve as the reader: across 26 readers, a stronger one narrows the gap to LALMs with no change to the audio side (\S\ref{sec:res-scaling}).
  \item \textbf{Interpretability by design.} Every answer is conditioned only on the nodes and transcript printed in the reader's prompt, and each prediction links to a traceable retrieval path through the tree (\S\ref{sec:reader}). A wrong answer can therefore be located in a node and corrected without touching the reader. The nodes are causal: replacing the deciding node with a distractor overturns 78\% of correct answers, and injecting the gold node repairs 91\% of wrong ones (\S\ref{sec:res-interp}).
  \item \textbf{Few-shot adaptation at low training cost.} A new domain is added as one attribute of the tree, with a frozen expert encoder and a small head trained on a few labelled clips; the reader and the other heads are not updated, so there is no catastrophic forgetting \citep{mccloskey1989catastrophic,kirkpatrick2017overcoming}: with both new domains added, accuracy on SAKURA, MMAU and MMAR drops by at most 0.6 points (\S\ref{sec:domain}, Appendix~\ref{app:domain}). With five clips per species, this outperforms QLoRA fine-tuning of an LALM on the same clips while training 120--240$\times$ fewer parameters.
\end{itemize}
We evaluate on MMAU, MMAR, SAKURA and MMAU-Pro, and on two new bioacoustic domains (\S\ref{sec:setup}, \S\ref{sec:results}).
\oursacc{} is a transparent alternative to end-to-end LALMs that is cheap to train: it trains 7.7M parameters, about 1{,}400$\times$ fewer than Qwen2.5-Omni-7B, on orders of magnitude less audio data (\S\ref{sec:res-cost}).
Demos, code and checkpoints, worked examples and an interactive viewer of the tree are available on our \href{\projecturl}{project page}.

%% file: sections/fig_pipeline.tex
\begin{figure}[t]
\centering
\includegraphics[width=0.79\linewidth]{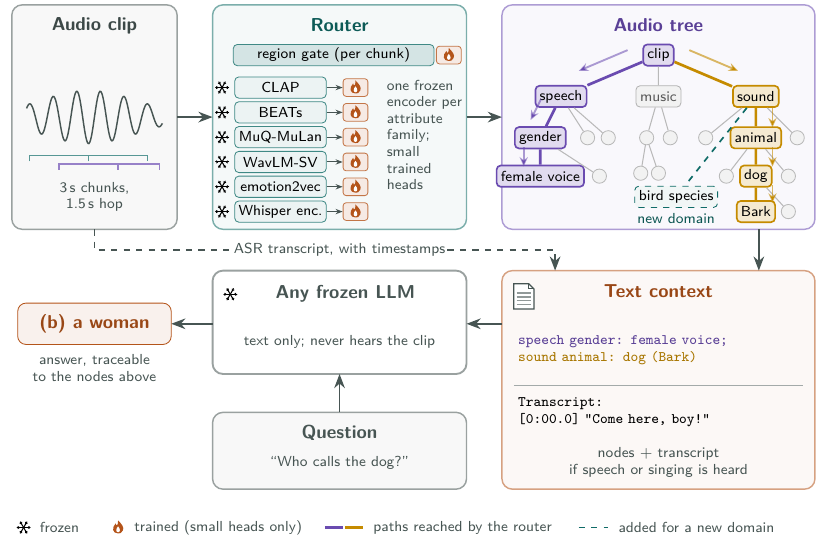}
\vspace{-.3cm}
\caption{\textbf{Overview of \ours{}.} \textbf{Listen:} a four-level tree describes speech, music and sound. \textbf{Retrieve:} router heads on frozen encoders map each chunk to tree nodes, which join an ASR transcript in a readable context. \textbf{Reason:} a frozen text-only LLM answers from this context. \textbf{Adapt:} a new domain is a new encoder--head branch (dashed).}
\label{fig:pipeline}
\vspace{-.4cm}
\end{figure}

%% file: sections/related.tex
\vspace{-.3cm}
\subsection{Related work}
\label{sec:related}

\vspace{-.2cm}
\paragraph{Audio LLMs and audio agents.}
LALMs range from early task-as-generation models \citep{deshmukh2023pengi,gong2024ltu} to multi-encoder and mixture-of-experts designs \citep{tang2024salmonn,shan2025pam,wang2025usam}; their encoder output enters the LLM as continuous embeddings, so adding a domain or changing the LLM requires further fine-tuning, typically with LoRA or QLoRA \citep{hu2022lora,dettmers2023qlora}.
Audio agents keep an LALM in the loop as reasoner or tool, and several train it to plan tool calls \citep{audiotoolagent2025,audiomaestro2025,wang2026audiomind,chen2026audiorouter,wang2026speechagentr}.
\ours{} also combines several frozen encoders, but their outputs reach the LLM only as named nodes in text; no LALM runs at inference, the same heads run on every clip without planning, and a new domain is a new head on a frozen encoder rather than a fine-tune, as in few-shot audio classification from prototypes or pretrained embeddings \citep{snell2017proto,kahl2021birdnet,ghani2023perch,rauch2024birdset,hagiwara2023beans}; we compare the two routes directly on the same clips.

\vspace{-.2cm}
\paragraph{Text interfaces, concept bottlenecks and retrieval.}
Cascades describe audio in text for a text LLM, through zero-shot model composition \citep{zeng2022socratic}, captions \citep{sakshi2024mmau,lu2026auditoryknowledge} or agent-refined descriptions \citep{rong2025audiogenie}.
SAR-LM \citep{taheri2025sarlm} writes the outputs of off-the-shelf extractors for speech, sound events and music as text, lets a frozen LLM reason over them, and traces errors to features.
\ours{} differs from it in four ways.
First, it describes every clip with one hierarchical vocabulary rather than the separate label spaces of several extractors.
Second, it predicts the nodes with small heads trained for this vocabulary, which back off from leaf to class when the leaf is uncertain, whereas SAR-LM uses its extractors as they are.
Third, it adds a new domain as one attribute with a head trained on a few labelled clips.
Fourth, we test its interpretability quantitatively: an audio LLM judge scores how related the selected nodes are to the clip, and interventions on the nodes test whether the answer depends on them.

Concept bottleneck models predict human-defined concepts before the label, so a person can inspect and correct them \citep{koh2020cbm,oikarinen2023lfcbm}, including for voice-based health assessment \citep{chen2026voicecbm}; in audio, post-hoc methods instead explain classifiers with listenable components or maps \citep{parekh2022l2i,paissan2024lmac,paissan2024lmaczs}.
Graph-based retrieval reaches knowledge from a text query \citep{edge2024graphrag,gutierrez2024hipporag} or from audio and multimodal keys \citep{chen2025wavrag,park2025m3kg,mousavi2025listen}.
Knowledge graphs have also been paired with audio models: iKnow-audio re-ranks the zero-shot predictions of CLAP with a graph of relations between sound events \citep{olvera2025iknowaudio}, and Audiopedia adds Wikidata facts about entities named in a clip's transcript to the prompt of an LALM \citep{penamakuri2025audiopedia}.
Such graphs store facts about sounds or entities, whereas our tree describes what is heard in a clip across speech, music and sound, and the reader supplies the world knowledge.
\ours{} is a concept bottleneck for open-ended audio question answering.
The waveform, not a text query, is the entry point; some heads are distilled from teachers that run only at training time \citep{hinton2015distill}; and the predictor above the bottleneck is any frozen LLM, which we vary across 26 readers.

%% file: sections/method.tex
\vspace{-.3cm}
\section{Method}
\label{sec:method}
\ours{} (Figure~\ref{fig:pipeline}) achieves interpretable, modular reasoning by constructing a hierarchical multimodal tree (\emph{Listen}), mapping input clips to tree nodes to generate human-readable context (\emph{Retrieve}), using a frozen LLM to answer queries based on this context (\emph{Reason}), and efficiently expanding to new domains by training lightweight projection heads (\emph{Adapt}).

\vspace{-.3cm}
\subsection{Listen: Building the Multimodal Tree}
\label{sec:graph}
We organize audio concepts into a four-level hierarchical tree: \emph{regions} (speech, music, environmental sound), \emph{attributes} characterizing each region (e.g., speaker gender, tempo, sound source), closed \emph{classes} under each attribute (e.g., \texttt{angry}, \texttt{jazz}, \texttt{engine}), and open-vocabulary \emph{leaves} that provide granular refinements (e.g., \texttt{engine knocking}). Domain knowledge informs the attributes, while environmental-sound identities align with the AudioSet ontology \citep{gemmeke2017audioset}.

To populate the nodes, we segment training clips into short chunks and annotate them using a hybrid approach. Each chunk is evaluated by single-task expert classifiers (e.g., speaker verification, emotion recognition) and a Large Audio Language Model. Rather than answering free-form queries, the LALM completes a strict template to fill predefined attributes: it identifies active regions, selects an existing class (or \emph{unsure}) for each attribute, and optionally appends a brief 1--3 word leaf (e.g., \texttt{\{is: happy, detail: sarcastic\}}). 

We merge the expert and LALM predictions depending on whether an attribute allows multiple concurrent choices (e.g., overlapping instruments or sound events) or only a single choice (e.g., gender or tempo). For attributes allowing multiple choices, we simply take the union of all predicted classes. For a single-choice attribute, we use the expert's label when its confidence exceeds a per-attribute threshold, and the LALM's label otherwise.
Finally, to maintain a clean taxonomy, free-text leaves are instantiated as nodes only if they recur across multiple clips. An external text LLM merges synonyms based on lexical and acoustic checks.
The annotation prompts, the full list of attributes and the leaf rules are given in Appendix~\ref{app:tree}.

\vspace{-.3cm}
\subsection{Retrieve: Routing Audio into the Tree}
\label{sec:router}

The router maps each short chunk $x$ of a clip to nodes of the tree.
Each attribute is read by a frozen pretrained encoder suited to it (e.g., CLAP for sounds and music, emotion2vec for emotion), and attributes that share an encoder form one \emph{router}: a small trainable body $\phi$ on the frozen embedding $h = E(x)$, with one head $g_a$ per attribute $a$.
A head has one output per node of its attribute and is trained to score highest the node that matches the chunk. For example, the emotion head has outputs for \texttt{happy}, \texttt{happy (sarcastic)}, \texttt{sad}, and so on; for a chunk spoken happily but sarcastically it should score \texttt{happy (sarcastic)} highest and \texttt{sad} low.
The score of a class $v$ combines its own node and its leaves $\mathcal{L}_{a,v}$, so \texttt{happy} is scored from both \texttt{happy} and \texttt{happy (sarcastic)}:
\begin{equation}
s_a(v \mid x) = \log \sum_{n \in \{v\} \cup \mathcal{L}_{a,v}} \exp g_a\big(\phi(h)\big)_n ,
\label{eq:lse}
\end{equation}
so a confident leaf also supports its class.
Each router is trained on all of its attributes at once. Since a chunk is labelled for only some attributes (e.g., gender but not language), the loss is masked,
\begin{equation}
\mathcal{L}(x) = \sum_{a} m_a(x) \Big[ \ell^{\mathrm{class}}_a(x) + \ell^{\mathrm{leaf}}_a(x) \Big],
\label{eq:loss}
\end{equation}
where $m_a(x) = 1$ only if attribute $a$ is labelled on $x$, and the two terms are cross-entropies over the classes and over the leaves of the labelled class.
At inference, a region gate decides which regions each chunk contains, each open attribute serves its most likely class, and a leaf is added only when it is confident; otherwise the router backs off to the class (e.g., \texttt{engine} instead of \texttt{engine knocking}).
Details are in Appendix~\ref{app:hparams}.

\vspace{-.3cm}
\subsection{Reason: Answering from the tree}
\label{sec:reason}
\label{sec:reader}

The router serves nodes for each chunk (\S\ref{sec:router}).
We collect them over the whole clip into one text context, listing each node once and grouping the nodes of each attribute together. For example, if a woman speaks in some chunks and a dog barks in others, the context reads \texttt{speech gender: female voice; sound animal: dog (Bark)}.
When speech or singing is heard, we add the ASR transcript, and for questions about the order, duration or number of events, we add the list of detected events with their times.
A frozen text LLM then receives this context with the question and answers it.
Because the reader only sees text, any LLM can be used without training, and each answer can be traced to, and changed by editing, the lines of its context.

\vspace{-.3cm}
\subsection{Adapt: Adding a domain}
\label{sec:adapt}
\label{sec:domain}

A new domain, such as bird species, becomes a new attribute under a region.
It needs a frozen encoder suited to the domain, either one of the existing encoders or a new one, and a head over that encoder trained on $k$ labelled clips per class.
We also train the head on clips without birds, so it learns to give them a low bird probability and the router does not enter the new domain for them.

%% file: sections/setup.tex
\vspace{-.3cm}
\section{Experimental details}
\label{sec:setup}
\label{sec:settings}

\vspace{-.2cm}
\paragraph{Data.}
\label{sec:data}
The tree is generated from 4{,}054 training clips from 19 sources, labelled by expert models and an audio LLM annotator (Qwen3-Omni-30B-A3B-Instruct \citep{xu2025qwen3omni}); the router heads also use human labels from AudioSet, FSD50K, ESC-50, MusicCaps, IRMAS and FLEURS (Appendix~\ref{app:data}). The tree has 39 attributes (speech 6, music 18, sound 15), 363 classes and 668 leaves (Appendix~\ref{app:tree}).

\vspace{-.2cm}
\paragraph{Benchmarks.}
\label{sec:benchmarks}
We evaluate on four multiple-choice audio reasoning benchmarks, scored by accuracy: MMAU test-mini \citep{sakshi2024mmau} (1{,}000 items over sound, music and speech), MMAR \citep{ma2025mmar} (1{,}000 items over speech, sound, music and their mixtures), SAKURA \citep{yang2025sakura} (4{,}000 single- and multi-hop items on gender, language, emotion and animal), and the single-audio multiple-choice subset of MMAU-Pro \citep{kumar2025mmaupro} (4{,}133 items).
For domain adaptation we build two species-identification tasks on which current LALMs are near chance: birds (48 species, 938 test items, from BirdSet \citep{rauch2024birdset}) and marine mammals (31 species, 678 test items, from BEANS \citep{hagiwara2023beans}). Each item is a multiple-choice question with four options, the correct species and three taxonomically close ones, so chance is 25\% (details in Appendix~\ref{app:benchmarks}).

\vspace{-.2cm}
\paragraph{Readers and baselines.}
\label{sec:baselines}
Because the router's output is cached as text, the same contexts are evaluated by 26 frozen text LLMs from 0.5B to 72B parameters, from the Qwen2.5, Qwen3, Llama, Vicuna, Mistral, Phi, OLMo, Falcon, SmolLM and gpt-oss families (Appendix~\ref{app:readers}).
Transcripts come from Whisper-large-v3 \citep{radford2023whisper}.
We compare against three LALMs that hear the clip: Qwen2.5-Omni-7B (Omni-7B) \citep{xu2025qwen25omni}, Audio Flamingo 3 (AF3) \citep{goel2025af3} and Qwen3-Omni-30B-A3B \citep{xu2025qwen3omni}.
Input ablations give a reader the question only (\emph{blind}), the transcript only, the tree only, or both.
For new domains, we compare our head with three uses of the same labelled clips by Omni-7B: zero-shot, clips given in context, and QLoRA fine-tuning \citep{dettmers2023qlora}.
Training and inference details are in Appendix~\ref{app:hparams}.

%% file: sections/results.tex
\vspace{-.3cm}
\section{Results}
\label{sec:results}

\vspace{-.3cm}
\subsection{Main evaluation on reasoning benchmarks}
\label{sec:res-main}

\begin{table}[t]
\centering
\caption{\textbf{Accuracy (\%) on four audio reasoning benchmarks}, full sets. Our readers never hear the clip; they read the same cached tree and transcript. B = parameters in billions (for audio LLMs, total size from the released weights). On MMAU-Pro the options are shuffled per item for all systems (Appendix~\ref{app:benchmarks}).}
\label{tab:main}
\small
\setlength{\tabcolsep}{5pt}
\resizebox{\linewidth}{!}{%
\begin{tabular}{llccccc}
\toprule
& & B & SAKURA$\uparrow$ & MMAU$\uparrow$ & MMAR$\uparrow$ & MMAU-Pro$\uparrow$ \\
\midrule
\multirow{3}{*}{\shortstack[l]{Input ablation\\(Qwen2.5-7B)}}
& Blind (question and options only) & 7.6 & 32.6 & 50.2 & 37.8 & 38.9 \\
& Transcript only & 7.6 & 51.8 & 60.1 & 47.8 & 44.6 \\
& Tree only & 7.6 & 74.7 & 57.5 & 40.3 & 39.3 \\
\midrule
\multirow{5}{*}{\shortstack[l]{\oursacc{} (ours)\\(tree + transcript)\\by reader}}
& Qwen2.5-7B & 7.6 & 80.1 & 68.3 & 51.6 & 45.7 \\
& Qwen2.5-14B & 14.7 & 83.3 & 69.4 & 55.6 & 48.5 \\
& Qwen2.5-32B & 32.8 & 85.1 & 71.4 & 55.7 & 49.8 \\
& Qwen2.5-72B & 72.7 & 85.3 & 70.5 & 56.6 & 50.1 \\
& Phi-4 & 14.7 & 83.5 & 71.7 & 54.9 & 46.1 \\
\midrule
\multirow{3}{*}{\shortstack[l]{Audio LLMs\\(hear the clip)}}
& Qwen2.5-Omni-7B & 10.7 & 73.3 & 75.1 & 63.6 & 57.4 \\
& Audio Flamingo 3 & 8.3 & 68.2 & 75.5 & 57.8 & 49.1 \\
& Qwen3-Omni-30B-A3B & 35.3 & 73.0 & 74.0 & 62.0 & 58.2 \\
\bottomrule
\end{tabular}}
\vspace{-.3cm}
\end{table}

Table~\ref{tab:main} compares \ours{}, read by five frozen text LLMs, with three audio LLMs that hear the clip.
With the Qwen2.5-7B reader, \ours{} outperforms every audio LLM on SAKURA (80.1 against 73.3 for the best, Omni-7B) and trails them on the other three benchmarks, by 6.8 points to Omni-7B on MMAU and by 12.0 on MMAR.
Note that these gaps come with a model whose every answer can be traced to the nodes it read, and which trains about 1{,}400$\times$ fewer parameters than Omni-7B on a training set orders of magnitude smaller (\S\ref{sec:res-interp}, \S\ref{sec:res-cost}).
Measured from a blind reader that sees only the question and options (50.2 on MMAU, 37.8 on MMAR), \ours{} recovers nearly three quarters of Omni-7B's gain on MMAU, about half on MMAR and about a third on MMAU-Pro, and more than all of it on SAKURA.

The input ablation shows what each part of the context contributes.
The tree carries the identity of what is heard: on SAKURA, whose questions ask for an animal, emotion, gender or language, the tree alone reaches 74.7 and adds 28.3 points over the transcript.
The transcript carries spoken content, which matters on MMAU and MMAR, where questions mix speech with sound and music.
The two channels are complementary: together they outperform either one alone on every benchmark.

Because the audio side is fixed, a stronger reader narrows these gaps without any retraining.
Reading the same cached contexts, Phi-4 closes half of the MMAU gap to Omni-7B (71.7), Qwen2.5-72B closes about 40\% of the MMAR gap (56.6), and the 32B and 72B readers pass AF3 on MMAU-Pro (49.8 and 50.1 against 49.1).
We study the reader in \S\ref{sec:res-scaling} and the question types behind the remaining gap in Appendix~\ref{app:lack}.

\vspace{-.3cm}
\subsection{Scaling the reader}
\label{sec:res-scaling}

\begin{figure}[!htb]
\centering
\includegraphics[width=0.78\linewidth]{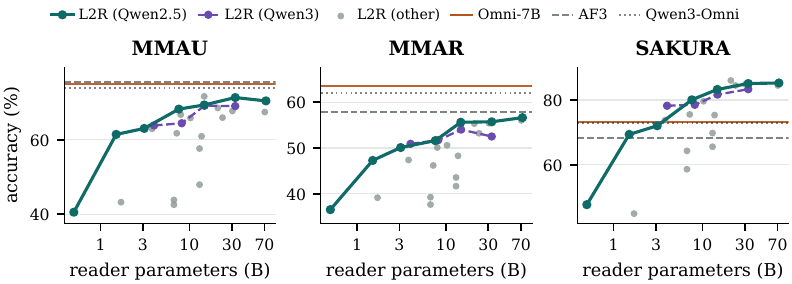}
\vspace{-.3cm}
\caption{\textbf{Accuracy of \ours{} as a function of the frozen reader.} Every reader reads the same cached tree and transcript contexts. Lines connect the Qwen2.5 (solid) and dense Qwen3 (dashed) families; grey points are the other 15 readers (Llama, Vicuna, Mistral, Phi, OLMo, Falcon, SmolLM, gpt-oss and the mixture-of-experts Qwen3-30B-A3B). Horizontal lines are audio LLMs that hear the clip. Full table in Appendix~\ref{app:readers}.}
\label{fig:scaling}
\vspace{-.1cm}
\end{figure}

The router and the ASR run once per clip and their output is cached, so any text LLM can read the same context, and differences between the 26 readers are differences in reasoning over a fixed description of the audio (Figure~\ref{fig:scaling}, Appendix~\ref{app:readers}).
Accuracy rises steeply with size up to 7B and more slowly beyond: within the Qwen2.5 family, the 7B reader already reaches 68.3 on MMAU and 80.1 on SAKURA, MMAR gains the most from 7B to 14B (4.0 points, to 55.6) because more of its questions require multi-step inference over the context, mostly the transcript, and above 14B all three benchmarks change by at most two points.

The generation of the reader matters as much as its size: Phi-4 (14.7B) is the best reader on MMAU (71.7), above Qwen2.5-72B (70.5), while Llama-2 and Vicuna of similar size stay close to their blind scores on MMAU and MMAR, barely using the identical context.
SAKURA shows how much the context alone carries: every reader released since 2024 with at least 7B parameters exceeds Omni-7B, which hears the clip. For single-hop questions the answer is usually the name of a node; for multi-hop questions the node name is enough because the reader's own knowledge supplies the rest: once the node identifies the animal as a dog, a follow-up question about dogs needs no further evidence from the audio.
Stronger readers can therefore be adopted as they appear, and each one reads the same cached description.

\vspace{-.3cm}
\subsection{Few-shot adaptation to new domains}
\label{sec:res-domain}

\begin{figure}[!htb]
\centering

\resizebox{0.75\linewidth}{!}{%
\begin{tikzpicture}
\begin{axis}[scale only axis, width=0.37\linewidth, height=0.16\linewidth, xmode=log, xmin=0.8, xmax=60, xtick={1,3,5,10,20,50}, xticklabels={1,3,5,10,20,50}, xlabel={labelled clips per species}, tick label style={font=\scriptsize}, label style={font=\scriptsize}, title style={font=\small\bfseries}, ymin=20, ymax=95, grid=major, grid style={gray!20}, every axis plot/.append style={line width=1pt}, legend style={font=\scriptsize, at={(1.08,-0.32)}, anchor=north, legend columns=4, draw=none, /tikz/every even column/.append style={column sep=6pt}}, title={Birds, 48 species}, ylabel={accuracy (\%)}]
\addplot[teal, mark=*] coordinates {(1,48.4)(3,72.7)(5,79.6)(10,85.3)(20,88.4)(50,91.1)};
\addplot[burnt, mark=square*] coordinates {(5,54.1)(10,61.0)(20,69.0)(50,75.2)};
\addplot[violet, mark=triangle*] coordinates {(1,43.3)(3,46.1)(5,47.3)};
\addplot[black, dashed, domain=0.8:60, samples=2] {32.3};
\legend{\oursacc{} head (ours), Omni-7B + QLoRA, Omni-7B in-context, Omni-7B zero-shot}
\end{axis}
\begin{axis}[scale only axis, at={(0.475\linewidth,0)}, width=0.37\linewidth, height=0.16\linewidth, xmode=log, xmin=0.8, xmax=60, xtick={1,3,5,10,20,50}, xticklabels={1,3,5,10,20,50}, xlabel={labelled clips per species}, tick label style={font=\scriptsize}, label style={font=\scriptsize}, title style={font=\small\bfseries}, ymin=20, ymax=95, grid=major, grid style={gray!20}, every axis plot/.append style={line width=1pt}, legend style={font=\scriptsize, at={(1.08,-0.32)}, anchor=north, legend columns=4, draw=none, /tikz/every even column/.append style={column sep=6pt}}, title={Marine mammals, 31 species}]
\addplot[teal, mark=*] coordinates {(1,41.3)(3,59.6)(5,65.2)(10,76.5)(20,83.0)};
\addplot[burnt, mark=square*] coordinates {(5,52.2)(10,68.6)(20,73.5)(50,83.6)};
\addplot[violet, mark=triangle*] coordinates {(1,44.1)(3,57.1)(5,60.5)};
\addplot[black, dashed, domain=0.8:60, samples=2] {25.8};
\end{axis}
\end{tikzpicture}}
\vspace{-.4cm}
\caption{\textbf{Data efficiency on new domains.} One small head (0.34M parameters for birds, 0.17M for marine mammals) on a frozen encoder (BirdNET for birds, CLAP for marine mammals), read by a frozen Qwen2.5-7B, against Omni-7B fine-tuned with QLoRA on the same clips and Omni-7B given the clips in context.}
\vspace{-.1cm}
\label{fig:domain}
\end{figure}
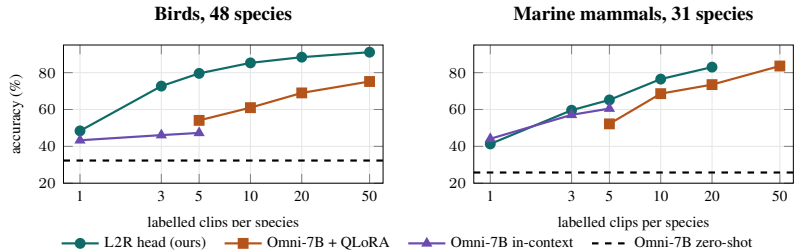

Next, we look at how \oursacc{} adapts to a new domain from a few labelled clips, a setting in which an LALM would normally have to be fine-tuned.
We use the two species-identification tasks of \S\ref{sec:benchmarks}, birds (48 species) and marine mammals (31 species), in which each question offers four candidate species. Both are out of distribution for current audio LLMs: Omni-7B scores 32.3 and 25.8, barely above chance (25).
\ours{} adds each domain by training one head on a frozen encoder suited to it: BirdNET \citep{kahl2021birdnet}, trained on bird vocalisations, for birds, and CLAP, already used by the router, for marine mammals. The heads have 0.34M and 0.17M parameters and train in under a minute on one GPU; the tree, the other heads and the reader stay unchanged.
We compare this head with two ways of giving the same clips to Omni-7B: fine-tuning it with QLoRA, and providing a few example clips with their labels in its prompt, known as in-context learning, which the LALM's context length limits to a few clips (at most five per candidate species here).

This head makes far better use of a few labelled clips than fine-tuning the LALM (Figure~\ref{fig:domain}).
With five clips per species, it reaches 79.6 on birds and 65.2 on marine mammals, against 54.1 and 52.2 for a QLoRA fine-tune of Omni-7B on the same clips, which trains 120--240$\times$ more parameters (40.4M) and takes 7--29 GPU-minutes per point.
The advantage persists as data grows: on birds, QLoRA with 50 clips per species (75.2) stays below our head with five, and on marine mammals it needs 50 clips to match what our head reaches with 20 (83.6 against 83.0).
Giving Omni-7B the clips in context helps less than QLoRA on birds and falls between the two methods on marine mammals, whose calls are short enough for the prompt to hold whole calls.

Because the domain node carries the answer, the reader barely matters: with the five-clip head, 24 readers from 1.5B to 72B all score 78--82 on birds and 64--68 on marine mammals.
Adding a domain also leaves the original tasks intact: with both heads added, the Qwen2.5-7B reader scores 79.6 on SAKURA (80.1 without), 67.7 on MMAU (68.3) and 51.8 on MMAR (51.6); exact numbers are in Appendix~\ref{app:domain}.

\vspace{-.3cm}
\subsection{Interpretability}
\label{sec:res-interp}

\begin{wrapfigure}{r}{0.45\linewidth}
\vspace{-10pt}
\centering
\includegraphics[width=\linewidth]{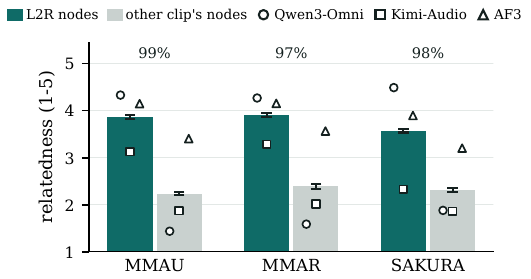}
\vspace{-14pt}
\caption{\textbf{Relatedness of the served nodes to the audio} (1--5). Bars: mean of three judges with 95\% bootstrap intervals; markers: each judge; numbers: share of items where the \oursacc{} nodes score higher.}
\label{fig:judge}
\vspace{-10pt}
\end{wrapfigure}
We now turn to interpretability.
Every answer of \ours{} is conditioned only on text that a person can read, the served nodes and the transcript, so the explanation of an answer is its own input.
For this explanation to be useful, the nodes must describe the audio correctly, and the answer must depend on them.
We test the first with audio LLMs as judges and the second with interventions on the nodes, and illustrate both with examples.
An interactive page plays example clips and shows the router's path through the tree, chunk by chunk, with the probability of every node and the text the reader receives.\footnote{\projectexamples}
\paragraph{Audio LLMs as judges.}
To check that the served nodes describe the audio, three audio LLMs (Qwen3-Omni-30B-A3B, Kimi-Audio-7B and Audio Flamingo 3) hear each clip and rate, on a 1--5 scale, how related the served nodes are to what is audible, without seeing the question.
They rate 1{,}200 held-out items (400 each from MMAU, MMAR and SAKURA, 100 per SAKURA track), and, as a control, the nodes of a clip from a different category (e.g.\ music nodes for a speech clip).
All three judges rate the served nodes well above the swapped ones on every benchmark, and higher on 97--99\% of items (Figure~\ref{fig:judge}).
They differ in scale, one rating even the swapped nodes above the midpoint, but they agree in direction; the prompt and protocol are in Appendix~\ref{app:judge}.
\vspace{-.4cm}
\paragraph{Case study.}
Figure~\ref{fig:examples} in Appendix~\ref{app:examples} shows what the reader actually receives for three MMAU items, and our project page shows many more similar examples.
In the two correct items, the node that decides the answer is present, and the reader picks the option that names it.
In the wrong item, the router reached the \texttt{water} class but served \texttt{Rain} instead of \texttt{Waterfall}, and the reader followed it, so the error can be traced to a single node, which is where a fix belongs.

\vspace{-.3cm}
\paragraph{The nodes are causal for the answer.}
The judges show that the nodes fit the audio; interventions show that the answer depends on them.
When we serve the tree of another clip together with the item's own transcript, Qwen2.5-7B drops from 68.3 to 55.0 on MMAU and from 51.6 to 46.4 on MMAR, below its transcript-only scores (60.1 and 47.8): the reader uses the nodes it is given, and a wrong tree is worse than none.
On SAKURA, with the tree of a clip from a different class, accuracy falls from 80.1 to 40.0; gender falls to 11.7\%, far below chance, because the reader follows the wrong gender node.
To isolate single nodes, we edit the context of SAKURA single-hop items, whose question asks for exactly one attribute, and change only that attribute's clause.
Replacing the routed node with the node of a distractor option breaks 78.3\% of the correct answers of Qwen2.5-7B, and replacing it with the gold node repairs 90.9\% of the wrong ones.
For animal, emotion and gender, the swap breaks 93.8\%, 96.9\% and 99.8\% of correct answers, and the inject repairs 87.5\%, 94.8\% and 97.8\% of wrong ones.
Language is the exception: the swap breaks only 8.0\%, because the transcript also shows which language is spoken, and the reader follows the words over a contradicting node.

\vspace{-.3cm}
\subsection{Efficiency and resource use}
\label{sec:res-cost}

Beyond accuracy, we compare what \ours{} and the audio LLMs cost (Table~\ref{tab:cost} in Appendix~\ref{app:cost}): training cost decides how easily a system can be rebuilt, extended or moved to a new LLM, and inference cost decides what it takes to deploy it.
Parameters are summed from the released weights, without the pretraining data of the components each system starts from; MACs (PyTorch FLOP counter) and seconds are averaged over 20 MMAU items on one A100 GPU.

\ours{} is far cheaper to train, in both parameters and data.
It trains 7.7M parameters, about 1{,}400$\times$ fewer than Omni-7B, and it learns from a training set orders of magnitude smaller than those of the audio LLMs: about 0.2M labelled clips in total.
The audio LLMs, in contrast, train every parameter on vast audio corpora: Omni-7B uses 300 billion tokens of audio-related data in its second pretraining stage alone \citep{xu2025qwen25omni}, Audio Flamingo 3 uses 9.6M question--answer pairs and 13.2M audio--caption pairs \citep{goel2025af3}, and Qwen3-Omni trains its audio encoder on 20 million hours of audio and its first pretraining stage on 0.77 trillion audio tokens \citep{xu2025qwen3omni}.
The saving carries over to extending the system: adding a domain updates 120--240$\times$ fewer parameters than QLoRA, and because the router's output is cached, reading the same audio with another LLM costs only that LLM's forward pass.

Inference, in contrast, is more expensive than for a dense audio LLM of similar size: per MMAU item, \ours{} costs 5.3 TMACs and 0.57\,s, against 2.7 TMACs and 0.12\,s for Omni-7B.
Most of this cost lies outside the tree: Whisper-large-v3 accounts for half of it (2.6T) and the reader's two passes, with and without the context, for 2.1T, while all six router encoders together take 0.6T; the encoders also run one after another, which adds to the time.
These figures compare a pipeline assembled from off-the-shelf components, run as they are, with audio LLMs built and tuned for inference as single networks.
The largest share of our cost, the ASR model, could be replaced by a smaller or distilled one, and the encoders could run in parallel.

%% file: sections/conclusion.tex
\vspace{-.3cm}
\section{Conclusion}
\label{sec:conclusion}
\ours{} replaces the fused audio representation of an audio language model with an explicit audio tree and a frozen text reader: small heads on frozen encoders map each chunk of a clip to named nodes, and a text LLM that never hears the clip answers from these nodes and a transcript.
Every answer can therefore be traced to the nodes it received, and audio LLMs judge those nodes as related to what is audible.
With a small reader, \ours{} outperforms audio LLMs on questions about the identity of a sound or speaker, and trails them with a relatively small gap on open reasoning over sound and music, despite being trained a small dataset, and having fewer trainable parameters.
The tree and the transcript are complementary, a stronger reader narrows the gap without changing any audio component, and a new domain is added with one small head and a few labelled clips, more effectively than by fine-tuning an audio LLM.
\vspace{-.2cm}
\paragraph{Limitations and future work.}
The tree has a closed vocabulary and describes short chunks rather than individual events, so it cannot yet express counts, the order of events, musical structure or the overall setting of a scene; these questions fall back on the transcript and the reader's prior, and they account for most of the remaining gap.
Event-level nodes and signal-level experts, such as music transcription, beat tracking and speaker diarization, can be added to the modular tree as new attributes.
The tree also inherits the errors of the annotators and teacher classifiers used to build it, and the reader follows a wrong node rather than correcting it; better heads and a reader that weighs node confidence are natural next steps.
Finally, inference runs several encoders and an ASR model before the reader, and domain addition has been tested only on identification tasks; lighter front ends and domains that require reasoning beyond identity remain to be explored.

%% file: sections/statements.tex
\section*{Ethics statement}
This work uses only publicly available datasets and models, under their licenses, and collects no new data from human participants; the two bioacoustic evaluation sets are drawn from BirdSet and the Watkins collection.
Some attributes of the tree describe speakers, such as gender, age group and emotion; such attributes can be misused to profile people, can be wrong for individuals, and follow the categories of their training data, which do not represent every speaker.
Because the tree was built partly from audio LLM annotations, it can also inherit their biases, although every answer is conditioned on readable nodes, so such errors can be inspected and corrected.
The healthcare applications that motivate this work were not evaluated, and \oursacc{} is not validated for clinical use.

\section*{Reproducibility statement}
The tree construction and annotation are described in \S\ref{sec:graph} and Appendix~\ref{app:tree}, the router, the heads and all hyperparameters in \S\ref{sec:router} and Appendix~\ref{app:hparams}, the datasets and splits in Appendix~\ref{app:data}, and the benchmarks, scoring and option shuffling in Appendix~\ref{app:benchmarks}.
Results for every reader and every number of labelled clips are in Appendices~\ref{app:readers} and~\ref{app:domain}, and the judge prompt and protocol in Appendix~\ref{app:judge}.
All encoders, readers and baselines are publicly released models, no benchmark audio is used for training, and random seeds are fixed for data splits, head training and template selection.
The code repository and all checkpoints needed to reproduce our results, including the tree and the trained router heads, are released on our \href{\projecturl}{project page}.

\section*{AI use statement}
Two components of the method use generative models, and both are described and evaluated in the paper: an audio LLM (Qwen3-Omni-30B-A3B-Instruct) annotates the training chunks from which the tree is built, and a text LLM (Llama-3.3-70B-Instruct) proposes leaf groups, which are accepted only after lexical, acoustic and identity checks (\S\ref{sec:graph}, Appendix~\ref{app:tree}); three audio LLMs also serve as judges of node relatedness (\S\ref{sec:res-interp}, Appendix~\ref{app:judge}).
In preparing the paper, we used an AI assistant to edit and draft passages of the text, suggest section structure, check descriptions of related work against the cited papers, and format bibliography entries.
We also used it for generic tasks, such as generating figures from our results and preparing reports of the experiments; we double-checked all such output and confirmed its validity.
The authors reviewed all AI-assisted text, checked every number against the experimental results and every citation against its source, and take full responsibility for the content.

%% file: sections/appendix.tex
\clearpage
\setcounter{topnumber}{3}\setcounter{bottomnumber}{2}\setcounter{totalnumber}{5}
\renewcommand{\topfraction}{0.9}\renewcommand{\bottomfraction}{0.8}\renewcommand{\textfraction}{0.07}\renewcommand{\floatpagefraction}{0.7}
\begin{center}
{\Large\sc Appendix}
\end{center}
\section{Tree construction details}
\label{app:tree}

This appendix follows the steps of \S\ref{sec:graph}: the attributes of the tree (\ref{app:tree-attrs}), the annotation of training chunks (\ref{app:annot}) and the processing of free-text leaves (\ref{app:tree-leaf}).

\subsection{Attributes}
\label{app:tree-attrs}

Table~\ref{tab:app-attrs} lists the attributes of each region, the number of classes and leaves, whether each attribute is decided once per clip or per chunk, and which expert classifier labels it.
Table~\ref{tab:app-sound} lists the identity attributes of environmental sound, whose classes follow the AudioSet ontology; their heads are trained on human labels from AudioSet, FSD50K and ESC-50, and the activity attribute also on the predictions of a VGGSound classifier.

\begin{table}[!htbp]
\centering
\caption{\textbf{Attributes of the tree.} * = multiple-choice attribute. Router encoder = frozen encoder whose embedding the attribute's head reads at training and inference. Expert labeller = expert classifier that labels the attribute when the tree is built, with its confidence threshold; it does not run at inference. Human labels = datasets whose labels are added to the training targets. $^{\dagger}$When a transcript is available, language is predicted from its text. New-domain attributes are added after the tree is built, each trained on $k$ labelled clips per species plus clips without that domain.}
\label{tab:app-attrs}
\small
\setlength{\tabcolsep}{3.5pt}
\resizebox{\linewidth}{!}{%
\begin{tabular}{llcccp{4.0cm}lll}
\toprule
Region & Attribute & Classes & Leaves & Decided per & Example classes & Router encoder & Expert labeller & Human labels \\
\midrule
all & regions present* & 3 & -- & chunk & speech, music, environmental sound & CLAP & PANNs (0.30) & -- \\
\midrule
\multirow{6}{*}{speech}
& gender & 2 & -- & chunk & male voice, female voice & WavLM-SV & WavLM-SV (0.05) & -- \\
& age group & 4 & -- & chunk & child, young adult, elderly adult & Whisper & -- & -- \\
& language & 7 & 50 & clip & western european (\texttt{french}), south asian & Whisper$^{\dagger}$ & Whisper-LID (0.50) & -- \\
& number of speakers & 3 & -- & clip & one, two, three or more & Whisper & -- & -- \\
& emotion & 7 & 50 & chunk & angry, happy, sad, neutral & emotion2vec & emotion2vec (0.60) & -- \\
& speaking style & 8 & 67 & chunk & conversational, public address, shouting & Whisper & -- & -- \\
\midrule
\multirow{18}{*}{music}
& instrument* & 22 & 71 & chunk & acoustic guitar, cello, drum kit & CLAP & -- & IRMAS, MusicCaps \\
& vocals & 6 & 27 & chunk & female vocals, choir, no vocals & CLAP & -- & MusicCaps \\
& vocal style* & 16 & -- & chunk & pop or rock singing, breathy, harmonized & CLAP & -- & MusicCaps \\
& ensemble size & 5 & -- & clip & solo instrument, full band, orchestra & CLAP & -- & -- \\
& genre & 19 & 31 & clip & blues, classical, electronic & CLAP & -- & -- \\
& mood & 10 & 42 & clip & calm, dark, energetic & CLAP & -- & MusicCaps \\
& tempo & 4 & -- & clip & slow, moderate, fast & MuQ-MuLan & librosa (0.30) & -- \\
& meter & 4 & -- & clip & duple, triple, irregular & MuQ-MuLan & -- & -- \\
& mode & 2 & -- & clip & major key, minor key & CLAP & librosa (0.80) & -- \\
& texture & 6 & 14 & clip & single melodic line, dense layered & MuQ-MuLan & -- & -- \\
& melody & 7 & 33 & chunk & ascending line, repetitive motif & MuQ-MuLan & -- & -- \\
& rhythm feel & 6 & -- & clip & syncopated, swung, march-like & MuQ-MuLan & -- & -- \\
& dynamics & 6 & -- & clip & steady, builds up, fades out & MuQ-MuLan & -- & -- \\
& form & 7 & -- & clip & verse and chorus, loop-based & MuQ-MuLan & -- & -- \\
& fidelity & 5 & -- & clip & clean, noisy, muffled recording & CLAP & -- & MusicCaps \\
& production & 5 & 24 & clip & studio, live performance, lo-fi & CLAP & -- & MusicCaps \\
& era / culture & 10 & 6 & clip & baroque, celtic, east asian & MuQ-MuLan & -- & -- \\
& period & 5 & -- & clip & contemporary popular, romantic era & MuQ-MuLan & -- & -- \\
\midrule
\multirow{3}{*}{sound}
& scene & 22 & 77 & clip & airport, farm, forest & CLAP & -- & -- \\
& action / material & 10 & 104 & chunk & footsteps, breaking, air flow & CLAP & -- & -- \\
& proximity & 4 & -- & chunk & close, at a distance, approaching & CLAP & -- & -- \\
\midrule
\multirow{2}{*}{\shortstack[l]{new domain\\(\S\ref{sec:res-domain})}}
& bird species & 48 & -- & clip & northern cardinal, American crow, hermit thrush & BirdNET & -- & BirdSet ($k$ per species) \\
& marine-mammal species & 31 & -- & clip & beluga, killer whale, bearded seal & CLAP & -- & Watkins ($k$ per species) \\
\bottomrule
\end{tabular}}
\end{table}

\begin{table}[!htbp]
\centering
\caption{\textbf{Identity attributes of environmental sound} (all multiple-choice), following the AudioSet ontology. Each attribute is read by CLAP or BEATs, whichever scores higher on held-out AudioSet and FSD50K clips. The new-domain heads of \S\ref{sec:res-domain} read BirdNET (birds) and CLAP (marine mammals).}
\label{tab:app-sound}
\small
\setlength{\tabcolsep}{4pt}
\begin{tabular}{lcclp{6.0cm}}
\toprule
Attribute & Classes & Leaves & Router encoder & Example classes \\
\midrule
animal & 26 & 96 & BEATs & bird, cat, cattle, dog, insect \\
vehicle & 12 & 40 & CLAP & aircraft, boat, car, emergency vehicle, train \\
machine & 2 & 37 & CLAP & engine, mechanisms \\
tool & 7 & 12 & CLAP & hammer, jackhammer, power tool, sanding \\
household & 23 & 38 & CLAP & blender, dishes, door, tap or sink, typing \\
human non-speech & 24 & 58 & BEATs & applause, breathing, cough, laughter \\
nature & 5 & 25 & CLAP & fire, thunderstorm, water, wind \\
alarm and signal & 13 & 34 & CLAP & alarm clock, beep, bell, siren, telephone \\
impact & 7 & 51 & CLAP & explosion or gunshot, glass, liquid, surface contact \\
environment & 8 & 10 & BEATs & echo, inside a small room, noisy background \\
sound type & 18 & 23 & CLAP & clang, clicking, creak, chanting \\
activity (VGGSound) & 10 & 34 & BEATs & cooking or eating, construction or repair, office work \\
\bottomrule
\end{tabular}
\end{table}

\subsection{Annotation}
\label{app:annot}

Each training chunk is labelled by the expert classifiers of Table~\ref{tab:app-attrs} and by the LALM (Qwen3-Omni-30B-A3B-Instruct).
The LALM hears the first 30\,s of a clip and fills the template in two steps: it first says which regions it hears, then answers one template per heard region, choosing a listed class, \emph{unsure} or \emph{none} for each attribute and optionally a short leaf.
Its system message is \emph{``You are a careful audio annotator. You describe only what can be heard, and you say unsure when you cannot tell.''}
The box shows the first step and the speech template, abbreviated to four attributes.

\begin{tcolorbox}[colback=gray!4, colframe=gray!60, boxrule=0.5pt, arc=2pt, left=4pt, right=4pt, top=2pt, bottom=2pt, fontupper=\scriptsize]
\textbf{Step 1 (regions):} Listen to the audio. Which of these are clearly audible? Answer with a JSON object only, e.g.\ \texttt{\{"speech": true, "music": false, "sound": true\}}. ``sound'' means environmental or object sounds other than speech and music. SINGING is part of music, not speech: set ``speech'' to true only when someone is TALKING.\\[4pt]
\textbf{Step 2 (form, one per present region):} Listen to the audio and fill in this form about the speech in it. For each field choose exactly one of the listed values (copy it verbatim), or ``unsure'' if you cannot tell from the audio, or ``none'' if the field does not apply. Do not guess. Where a field is an object, ``is'' is the listed value and ``detail'' is the SINGLE most specific term (1--3 plain words of YOUR OWN, no ``and'') for what you hear under that value. No numbers in a detail. An empty detail \texttt{""} is the RIGHT answer whenever you can add nothing specific.\\
-- \texttt{"gender"}: male voice $|$ female voice\\
-- \texttt{"emotion"}: \texttt{\{"is":} one of angry $|$ happy $|$ sad $|$ neutral $|$ fearful $|$ disgusted $|$ surprised, \texttt{"detail": "..."\}}\\
-- \texttt{"speaking\_style"}: \texttt{\{"is":} one of conversational speech $|$ read speech $|$ public address $|$ shouting $|$ \ldots, \texttt{"detail": "..."\}}\\
-- \texttt{"language"}: \texttt{\{"is":} one of western european $|$ south asian $|$ \ldots, \texttt{"detail": "..."\}} -- detail = the name of the language, one word\\
-- \texttt{"notable"}: one short sentence about anything important in the audio that this form does not cover, or \texttt{""} if nothing.\\
Answer with one JSON object only, no other text.
\end{tcolorbox}

For a single-choice attribute, the expert's label is used when its confidence is above its threshold and the LALM's label otherwise; for a multiple-choice attribute, the labels of both are combined.
Answers outside the listed classes, \emph{unsure} and \emph{none} give no label, and the \texttt{detail} answers are the candidate leaves.

\subsection{Leaf post-processing}
\label{app:tree-leaf}

Candidate leaves become nodes in four steps (Table~\ref{tab:app-leaf}).
The merging step uses an external text LLM (Llama-3.3-70B-Instruct), which proposes groups of names but never hears audio; its acoustic check compares the chunks of each proposed group in the embedding space of an encoder suited to the attribute.

\begin{table}[!htbp]
\centering
\caption{\textbf{Leaf post-processing.}}
\label{tab:app-leaf}
\small
\setlength{\tabcolsep}{4pt}
\begin{tabular}{lp{10.6cm}}
\toprule
Step & Rule \\
\midrule
Normalize & Remove filler words (\texttt{a}, \texttt{some}, \texttt{sound}) and merge names that are nearly identical and have the same number of words. \\
Admit & Keep a leaf only if it appears in at least 2 clips from at least 2 sources, adds detail beyond its class, and contains no number. \\
Merge & The text LLM proposes groups of synonyms. A group is kept only if its names do not conflict in number, negation or polarity (\texttt{high-pitched} vs \texttt{low-pitched}), its chunks sound closer to each other than to the other leaves of the class, and it does not join two different sources (\texttt{goat bleating} vs \texttt{sheep bleating}). \\
Prune & Fold a leaf that only adds a filler word into its sibling (\texttt{alarm sound} $\to$ \texttt{alarm}). Remove attributes whose content lies in the spoken words (topic, intent, scenario, speaker role) or that carried no usable information (recording channel, voice quality, accent, weather, number of sources). \\
\bottomrule
\end{tabular}
\end{table}

\FloatBarrier
\section{Implementation details}
\label{app:hparams}

This section follows the pipeline: router training, router inference, the reader, compute, and finally the QLoRA baseline.

\subsection{Router training}
\label{app:router}

All encoders are frozen, and their chunk embeddings are computed once and cached.
Let $y$ be the labelled class of attribute $a$ on chunk $x$.
For a single-choice attribute, class probabilities are a softmax over the class scores of Eq.~\ref{eq:lse}, $p_a(v \mid x) = \exp s_a(v \mid x) / \sum_{v'} \exp s_a(v' \mid x)$, and the class term of Eq.~\ref{eq:loss} is $\ell^{\mathrm{class}}_a = -\log p_a(y \mid x)$; for a multiple-choice attribute, $p_a(v \mid x) = \sigma\big(s_a(v \mid x)\big)$ and the class term is the binary cross-entropy against the set of labelled classes.
When the label also names a leaf $\ell^{*}$ under $y$, the leaf term compares only the nodes inside $y$,
\begin{equation}
\ell^{\mathrm{leaf}}_a = -\log \frac{\exp g_a\big(\phi(h)\big)_{\ell^{*}}}{\sum_{n \in \{y\} \cup \mathcal{L}_{a,y}} \exp g_a\big(\phi(h)\big)_n },
\label{eq:leaf}
\end{equation}
so learning a leaf never moves probability between classes.
A leaf becomes a node only with at least 5 training clips (3 for emotion and gender); a rarer leaf trains toward its class node, so the head learns when not to be specific.
Teacher labels (e.g.\ activity from a VGGSound classifier) are summed onto the tree nodes they map to and used as soft targets.
The environmental-sound identity heads predict three levels, the identity attribute $a$ (e.g.\ \texttt{animal}), the class $v$ (\texttt{dog}) and the leaf $\ell$ (\texttt{Bark}), each trained only where its parent is labelled, and serve $p(\ell \mid x) = p(a \mid x)\,p(v \mid a, x)\,p(\ell \mid v, x)$.
Table~\ref{tab:heads} gives the settings of every head.

\begin{table}[!htbp]
\centering
\caption{\textbf{Head settings.} All heads use AdamW; checkpoints are selected on held-out training data.}
\label{tab:heads}
\small
\setlength{\tabcolsep}{3pt}
\resizebox{\linewidth}{!}{%
\begin{tabular}{llllll}
\toprule
Heads & Architecture & Loss & Learning rate & Batch & Training \\
\midrule
speech, music, scene & 1--3-layer MLP body (width 256), one head per attribute & masked CE / BCE (Eq.~\ref{eq:loss}) & $3\!\times\!10^{-4}$--$10^{-3}$, one-cycle & 64 & 60--80 epochs \\
environmental-sound identity & LayerNorm + 1{,}024-unit layer, sigmoid per node & masked BCE, positive weights & $10^{-3}$ & 512 & 25 epochs \\
music properties & 2-layer MLP (width 256) & class-weighted CE & $10^{-3}$ & 64 & 60 epochs \\
event timeline & linear layer on PANNs frames & BCE on strong labels & $10^{-4}$ & 256 & 6 epochs \\
new domain & 1-layer MLP body, classes + \texttt{none} & CE & $10^{-3}$ & 64 & 40 epochs \\
\bottomrule
\end{tabular}}
\end{table}

Attributes are sampled in proportion to the square root of their number of labels, and part of each batch is class-balanced.
Leaf thresholds are fitted for 0.75 precision, and the thresholds of the sound identity heads for F1, on held-out training data.

\subsection{Router inference}

\begin{itemize}[leftmargin=*,itemsep=1pt]
  \item Clips are cut into 3\,s chunks with a 1.5\,s hop; a chunk reads only the regions its gate opens (probability $\geq 0.5$).
  \item Clip-level attributes average their log-probabilities over the chunks; chunk-level attributes (Table~\ref{tab:app-attrs}) decide per chunk.
  \item A single-choice attribute serves its best class (emotion: up to three classes with probability $\geq 0.2$); multiple-choice music attributes serve every class above 0.3, and each sound identity attribute its top two classes and leaves.
  \item Attributes with low held-out skill (balanced accuracy $< 0.3$ above chance) are served only at confidence $\geq 0.5$.
  \item If no chunk hears environmental sound, a question classifier (logistic regression on MiniLM embeddings) decides whether the answer names a sound; if so, the sound identity heads read the whole clip.
  \item The transcript (Whisper-large-v3, first 180\,s) is added when speech or singing is heard, and language is then predicted from its text; the event timeline is added only for questions about the order, duration or number of events.
  \item In the context, nodes appear in the order they were first reached; time spans and probabilities stay in the traversal trace.
  \item A new-domain head serves a class only when $p(\texttt{none}) < 0.995$; its \texttt{none} class is trained on 2{,}000 chunks of the training pool.
\end{itemize}

\subsection{Reader}

The reader does not generate free text for the answer. With context $c$, question and options $q$ and option letters $\mathcal{O}$, the prediction is
\begin{equation}
\hat{o} = \arg\max_{o \in \mathcal{O}} \Big[\log p(o \mid c, q) + \beta \log p(o \mid q)\Big],
\label{eq:answer}
\end{equation}
where the second term is the same reader without the context and $\beta = 0.1$; this removes answer-parsing errors and adds a weak prior from the reader's own knowledge.
The 26 readers are Qwen2.5 (0.5B--72B), Qwen3 (4B--32B, and 30B-A3B), Llama-2-chat (7B, 13B), Llama-3.1-8B, Llama-3.3-70B, Vicuna v1.5 (7B, 13B), Mistral-7B v0.3, Mistral-Small-24B, Phi-3.5-mini, Phi-4, OLMo-2-13B, Falcon3-10B, SmolLM2-1.7B and gpt-oss-20b.
For each reader, one of four prompt templates is selected on 100 development items per benchmark and then locked; gpt-oss, which reasons before answering, generates up to 512 tokens and its answer is parsed.

\subsection{Compute}

All experiments run on single NVIDIA A100 (40 or 80\,GB) or L40S GPUs; the tree annotation and the LLM cleaning use up to four A100-80GB GPUs.
A router head trains in a few minutes and a new-domain head in under one minute; a 7B reader reads all three benchmarks in about 5 minutes.
The 70B and 72B readers run in 4-bit NF4 quantization with bfloat16 computation, all others in bfloat16.
We use PyTorch 2.12, Transformers 4.57, PEFT 0.19, bitsandbytes 0.50, scikit-learn 1.7 and sentence-transformers 5.6, with fixed random seeds.

\subsection{QLoRA baseline}

The thinker of Omni-7B is loaded in 4-bit NF4 with LoRA adapters of rank 16 ($\alpha = 32$, dropout 0.05) on the attention and MLP projections of the language model (40.4M trainable parameters); the audio encoder is frozen.
Training uses AdamW with learning rate $2 \times 10^{-4}$, batch size 4 with 4 gradient-accumulation steps, a cosine schedule with 5\% warm-up, and 8, 6, 4 and 2 epochs for $k = 5, 10, 20, 50$ (3 epochs at $k = 50$ on marine mammals), on items in the same format as the evaluation items.

\FloatBarrier
\section{Datasets}
\label{app:data}

Table~\ref{tab:data} lists every dataset used to build the tree and train the router, with its role and size.

\begin{table}[!htbp]
\centering\small
\caption{\textbf{Datasets} used to generate the tree and to train the router. Tree pool = clips in the tree-generation pool (4{,}054 in total, labelled by the annotators). Router = additional labelled data used to train heads; ``--'' means the dataset reaches the router only through the pool's labels (the speech heads, including emotion, are trained on the pool alone). AudioSet test (9{,}042 clips) and FSD50K evaluation (10{,}161 clips) are held out. No benchmark audio is used.}
\label{tab:data}
\setlength{\tabcolsep}{4pt}
\begin{tabular}{@{}llr p{5.3cm}@{}}
\toprule
Dataset & Domain & Tree pool & Router training (labels, size) \\
\midrule
WavCaps AudioSet-SL & sound & 805 & sound hierarchy: AudioSet labels, 81{,}325 clips \\
Clotho & sound & 645 & -- \\
FSD50K & sound & 413 & sound hierarchy (CLAP heads): leaf labels, 24{,}238 dev clips \\
ESC-50 & sound & 280 & sound hierarchy (CLAP heads): class labels, 1{,}793 clips \\
WavCaps SoundBible & sound & 127 & -- \\
AudioCaps & sound & 107 & -- \\
Freesound & sound & 82 & -- \\
YouTube (AF-Think clips) & mixed & 82 & -- \\
AF-Think question--answer records & text & -- & question-path classifier: 83{,}249 records \\
AudioSet strong labels & sound & -- & event timeline: frame labels, 81{,}325 clips (the same split) \\
VGGSound (ECAPA classifier) & sound & -- & activity: soft targets (teacher, training only) \\
\midrule
MusicCaps & music & 618 & music heads: descriptors \\
IRMAS & music & 80 & music heads: instrument labels \\
FMA & music & 82 & -- \\
GTZAN & music & 80 & -- \\
Instrument samples & music & 22 & -- \\
\multicolumn{4}{l}{\quad MusicCaps + IRMAS for the music heads: 8{,}474 chunks} \\
\midrule
FLEURS & speech & 204 & language classifier: transcripts, 49 languages \\
RAVDESS & speech & 165 & -- \\
IEMOCAP & speech & 152 & -- \\
LibriSpeech & speech & 30 & -- \\
\midrule
TAU Urban Acoustic Scenes & scenes & 80 & -- \\
\midrule
BirdSet POW & birds & -- & domain head: $k$ clips per species \\
Watkins (BEANS) & marine & -- & domain head: $k$ clips per species \\
\bottomrule
\end{tabular}
\end{table}

\FloatBarrier
\section{Benchmarks and evaluation protocol}
\label{app:benchmarks}

This section details the benchmarks summarised in \S\ref{sec:benchmarks}, how items are scored, and the baseline settings.

\paragraph{MMAU.} Test-mini split, 1{,}000 items, about one third per domain, 27 skills that require expert knowledge.
\paragraph{MMAR.} 1{,}000 items in four reasoning layers (signal, perception, semantic, cultural) and 16 sub-categories. The answer is given as option text, which we match to the option letter (Appendix~\ref{app:gold}); three items whose answer names two options are removed.
\paragraph{SAKURA.} Four attributes (gender, language, emotion, animal), each with 500 single-hop and 500 multi-hop items. Single-hop items ask for the attribute; multi-hop items require the attribute and a fact about it.
\paragraph{MMAU-Pro.} Extends MMAU with long recordings, multi-audio questions, spatial audio, open-ended answers and 49 skills. We keep the single-audio multiple-choice items (4{,}133, with 2 to 11 options); multi-audio, open-ended and instruction-following items are not supported by our pipeline. For recordings longer than 3 minutes, the router covers the first 60\,s and the transcript the first 180\,s. The gold letters are strongly unbalanced (97\% of the spatial-audio answers are option (a)), so we shuffle the options of every item with a fixed per-item permutation, the same for all systems.
\paragraph{Scored items.} Three items per benchmark serve as in-context examples when a reader's prompt template is selected and are excluded from scoring, which leaves 997 MMAU, 994 MMAR and 3{,}997 SAKURA items.
\paragraph{New domains.} Birds: 938 items over 48 species from the Powdermill (POW) subset of BirdSet, with recordists disjoint from the training clips. Marine mammals: 678 items over 31 species from the Watkins collection, through BEANS. Each item has four options, and distractors are taxonomically close species (same genus first), so the option list alone does not reveal the answer. Training clips come from the same collections, excluding the evaluation recordings and any clip above 0.95 CLAP similarity to an evaluation clip.
\paragraph{Baselines.} Omni-7B is run with the same questions, options and scoring as our system, and hears the first 180\,s of each MMAU-Pro recording. AF3 is trained with the AF-Think chain-of-thought data. In the in-context baseline, $S \in \{1, 3, 5\}$ labelled clips of each of the item's four option species are given in the prompt; QLoRA uses the same $k$ clips per species with the same containment filter.
\paragraph{Scoring.}\label{app:gold} MMAR's \texttt{answer} field is the option text. We map it to an option letter by matching the text to the options (995 items exact, 2 by unambiguous word overlap). Three items whose answer names two options are removed. All systems, including Omni-7B, are scored with the same mapping. On MMAU the gold letter agrees with the option text on all 1{,}000 items, but 16 items list the correct answer text under two letters and score only one of them; counting either letter changes Omni-7B by $+0.6$ and ours by $+0.3$ points. We report the official scoring.

\FloatBarrier
\section{Which attributes the reader uses}
\label{app:ablation}

To decide which attributes to serve, we used the annotator's own labels on the 300 development clips per benchmark as a perfect router and removed one attribute group at a time from tree + transcript (Table~\ref{tab:attrgroups}, Qwen2.5-7B, before the pruning of Appendix~\ref{app:tree}).
Identity attributes (gender, sound source, emotion and the leaves) carry SAKURA, and music attributes carry MMAU; no group moves MMAR by more than 1.3 points (4 items).
Removing the speech content attributes (topic, scenario, speaker role, intent) does not lower SAKURA or MMAU and costs 1.3 points on MMAR; we prune them because their content is already in the transcript.

\begin{table}[!htbp]
\centering
\caption{\textbf{Attribute groups.} Change in accuracy (points) when one attribute group is removed from a perfect-router tree, on the development clips (SAKURA 600 items, MMAU 300, MMAR 298; one item is 0.2, 0.3 and 0.3 points). Negative values mean the group helps.}
\label{tab:attrgroups}
\small
\setlength{\tabcolsep}{5pt}
\begin{tabular}{lccc}
\toprule
removed group & SAKURA & MMAU & MMAR \\
\midrule
every leaf (classes kept) & $-8.0$ & $+0.3$ & $-1.3$ \\
speech gender & $-7.7$ & $-1.0$ & $+0.3$ \\
sound source (event, action and material) & $-7.2$ & $-1.3$ & $-0.3$ \\
speech emotion & $-7.0$ & $+1.3$ & $-0.7$ \\
speech language & $-0.8$ & $+1.0$ & $+0.7$ \\
music identity (genre, instrument, mood, era, vocals) & $0.0$ & $-2.3$ & $+0.3$ \\
music structure (tempo, meter, mode, texture, melody, production, ensemble) & $0.0$ & $-1.3$ & $0.0$ \\
speech voice (age, style, channel, voice quality, accent, number of speakers) & $-0.2$ & $0.0$ & $0.0$ \\
sound context (scene, activity, weather) & $-0.2$ & $+1.7$ & $-0.7$ \\
sound proximity and number of sources & $0.0$ & $+0.3$ & $0.0$ \\
speech content (topic, scenario, speaker role, intent) & $+0.5$ & $+0.7$ & $-1.3$ \\
\bottomrule
\end{tabular}
\end{table}

\FloatBarrier
\section{Choice of reader}
\label{app:readers}

Table~\ref{tab:readers} gives the accuracy of all 26 frozen readers on the same cached contexts; Figure~\ref{fig:scaling} plots them against reader size.

\begin{table}[!htbp]
\centering
\caption{\textbf{Choice of reader.} Accuracy (\%) of every frozen reader on the same cached tree + transcript contexts, full sets. B = parameters in billions (Qwen3-30B-A3B has 3B active). Prompt template locked per reader on 100 development items per benchmark. Llama-2 and Vicuna are scored with an answer prefix; gpt-oss by generation. Audio LLMs: B is total size from the released weights.}
\label{tab:readers}
\small
\begin{tabular}{lrccc}
\toprule
Reader & B & MMAU$\uparrow$ & MMAR$\uparrow$ & SAKURA$\uparrow$ \\
\midrule
Qwen2.5-0.5B & 0.5 & 40.5 & 36.5 & 47.6 \\
Qwen2.5-1.5B & 1.5 & 61.5 & 47.3 & 69.3 \\
SmolLM2-1.7B & 1.7 & 43.2 & 39.1 & 44.9 \\
Qwen2.5-3B & 3 & 63.1 & 50.1 & 72.0 \\
Phi-3.5-mini & 3.8 & 63.0 & 47.4 & 73.8 \\
Qwen3-4B & 4 & 63.9 & 50.9 & 78.2 \\
Llama-2-7B-chat & 6.7 & 43.8 & 39.2 & 58.6 \\
Vicuna-7B v1.5 & 6.7 & 42.5 & 37.6 & 64.3 \\
Mistral-7B v0.3 & 7.2 & 61.8 & 46.2 & 75.5 \\
Qwen2.5-7B & 7.6 & 68.3 & 51.6 & 80.1 \\
Llama-3.1-8B & 8 & 66.8 & 50.1 & 79.5 \\
Qwen3-8B & 8.2 & 64.5 & 51.7 & 78.5 \\
Falcon3-10B & 10.3 & 65.9 & 50.6 & 79.6 \\
Llama-2-13B-chat & 13 & 47.9 & 43.6 & 65.5 \\
Vicuna-13B v1.5 & 13 & 57.7 & 41.6 & 69.8 \\
OLMo-2-13B & 13.7 & 61.0 & 48.3 & 75.3 \\
Qwen2.5-14B & 14.7 & 69.4 & 55.6 & 83.3 \\
Phi-4 & 14.7 & \textbf{71.7} & 54.9 & 83.5 \\
Qwen3-14B & 14.8 & 69.2 & 54.0 & 81.7 \\
gpt-oss-20b & 21 & 68.6 & 55.3 & \textbf{86.0} \\
Mistral-Small-24B & 24 & 66.0 & 53.2 & 84.5 \\
Qwen3-30B-A3B & 30.5 & 67.8 & 55.4 & 84.9 \\
Qwen2.5-32B & 32.8 & 71.4 & 55.7 & 85.1 \\
Qwen3-32B & 32.8 & 69.1 & 52.5 & 83.3 \\
Llama-3.3-70B (4-bit) & 70.6 & 67.5 & 56.0 & 84.5 \\
Qwen2.5-72B (4-bit) & 72.7 & 70.5 & \textbf{56.6} & 85.3 \\
\midrule
Qwen2.5-Omni-7B (hears) & 10.7 & 75.1 & 63.6 & 73.3 \\
Audio Flamingo 3 (hears) & 8.3 & 75.5 & 57.8 & 68.2 \\
Qwen3-Omni-30B-A3B (hears) & 35.3 & 74.0 & 62.0 & 73.0 \\
\bottomrule
\end{tabular}
\end{table}

\FloatBarrier
\section{New domains: exact numbers}
\label{app:domain}

Table~\ref{tab:domain} gives the accuracy behind Figure~\ref{fig:domain} for every number of labelled clips per species.

\begin{table}[!htbp]
\centering
\caption{\textbf{Few-shot domain adaptation}, 4-way species identification, accuracy (\%) by labelled clips per species ($k$). Ours = one new head on a frozen encoder, frozen Qwen2.5-7B reader, tree only. Omni-7B zero-shot: birds 32.3, marine 25.8. Marine has a median of 33 labelled clips per species, so $k{=}50$ is not reachable for our head.}
\label{tab:domain}
\small
\setlength{\tabcolsep}{4.5pt}
\begin{tabular}{llcccccc}
\toprule
& method (accuracy$\uparrow$) & $k{=}1$ & 3 & 5 & 10 & 20 & 50 \\
\midrule
\multirow{3}{*}{Birds (48 sp.)} & \textbf{\oursacc{} (ours, 0.34M trainable)} & \textbf{48.4} & \textbf{72.7} & \textbf{79.6} & \textbf{85.3} & \textbf{88.4} & \textbf{91.1} \\
& Omni-7B + QLoRA (40.4M trainable) & -- & -- & 54.1 & 61.0 & 69.0 & 75.2 \\
& Omni-7B, clips in context & 43.3 & 46.1 & 47.3 & -- & -- & -- \\
\midrule
\multirow{3}{*}{Marine (31 sp.)} & \textbf{\oursacc{} (ours, 0.17M trainable)} & 41.3 & \textbf{59.6} & \textbf{65.2} & \textbf{76.5} & \textbf{83.0} & -- \\
& Omni-7B + QLoRA (40.4M trainable) & -- & -- & 52.2 & 68.6 & 73.5 & 83.6 \\
& Omni-7B, clips in context & \textbf{44.1} & 57.1 & 60.5 & -- & -- & -- \\
\bottomrule
\end{tabular}
\end{table}

\paragraph{The encoder matters, the reader does not.}
With the $k{=}5$ head, 24 readers from 1.5B to 72B all score 78--82 on birds and 64--68 on marine mammals.
The domain node carries the answer, so the reader only needs to copy it.
The choice of frozen encoder, in contrast, decides the result.
We use BirdNET for birds because it is trained on bird vocalisations, while CLAP is trained on general audio; for marine mammals we use CLAP, which the router already runs, so this domain adds no new encoder.
A modular design lets each domain use the encoder that hears it best.
\paragraph{Forgetting.} Adding both domain heads leaves the other heads unchanged, but the new nodes can be served on clips of the original benchmarks.
With both heads added, the Qwen2.5-7B reader scores 79.6 on SAKURA (80.1 without), 67.7 on MMAU (68.3) and 51.8 on MMAR (51.6).
A species node is served on 17--21\% of the items. All 19 SAKURA items it breaks are animal clips that now also carry a bird species (a cat served as \emph{black-and-white warbler}).
A tighter abstention rule for the domain heads is the fix.

\FloatBarrier
\section{Where \ours{} lags behind audio LLMs}
\label{app:lack}

\subsection{By question type}

Table~\ref{tab:lack} breaks the gap to Omni-7B down by question type, assigned from the question wording.

\begin{table}[!htbp]
\centering
\caption{\textbf{Where \ours{} lacks}, by question type (assigned from the question wording), against Omni-7B and the blind reader, scored against gold. Net = items we get right minus items Omni gets right. Qwen2.5-7B reader.}
\label{tab:lack}
\scriptsize
\setlength{\tabcolsep}{3.2pt}
\begin{tabular}{l rrrrr rrrrr}
\toprule
& \multicolumn{5}{c}{MMAU} & \multicolumn{5}{c}{MMAR} \\
\cmidrule(lr){2-6}\cmidrule(lr){7-11}
question type & n & ours$\uparrow$ & Omni$\uparrow$ & blind$\uparrow$ & net$\uparrow$ & n & ours$\uparrow$ & Omni$\uparrow$ & blind$\uparrow$ & net$\uparrow$ \\
\midrule
counting & 100 & 47.0 & 53.0 & 43.0 & -6 & 122 & 33.6 & 45.1 & 27.0 & -14 \\
order / timing & 113 & 60.2 & 62.8 & 39.8 & -3 & 106 & 47.2 & 57.5 & 37.7 & -11 \\
scene / place / activity & 115 & 77.4 & 93.0 & 46.1 & -18 & 111 & 58.6 & 71.2 & 27.0 & -14 \\
identify source / instrument / genre & 148 & 71.6 & 83.8 & 60.1 & -18 & 30 & 43.3 & 46.7 & 40.0 & -1 \\
music theory & 102 & 57.8 & 61.8 & 52.9 & -4 & 22 & 50.0 & 31.8 & 40.9 & +4 \\
speaker / voice & 202 & 74.8 & 72.3 & 48.0 & +5 & 74 & 56.8 & 77.0 & 31.1 & -15 \\
speech content & 37 & 89.2 & 100.0 & 43.2 & -4 & 82 & 58.5 & 65.9 & 46.3 & -6 \\
other & 180 & 71.1 & 81.1 & 52.2 & -18 & 447 & 54.4 & 68.0 & 40.0 & -61 \\
\bottomrule
\end{tabular}
\end{table}

The losses fall into three groups.

\paragraph{No node carries the answer: counting.}
On counting questions \ours{} stays close to the blind reader (MMAU 47.0 vs 43.0; MMAR 33.6 vs 27.0), a smaller gain over blind than on any other type.
The router describes 3\,s chunks and does not represent individual event occurrences, so no node states a count.
Omni-7B is also weak here (53.0 and 45.1), which suggests that counting is hard for current audio encoders in general.
Order and timing questions are closer: \ours{} is within three points of Omni-7B on MMAU (60.2 vs 62.8), though 11 items behind on MMAR.

\paragraph{A node exists but is too coarse: identity and scene.}
The largest MMAU losses are in identifying a source, instrument or genre ($-18$ items) and in describing a scene or activity ($-18$).
The answer to these questions is a specific name (\emph{waterfall}, not \emph{water}) or a description of the whole clip.
In the errors we inspected, the router often reaches the right class but not the right leaf, or serves several candidate leaves.
Figure~\ref{fig:examples} shows one such case.

\paragraph{The answer is in the transcript but needs inference: MMAR.}
On MMAR the largest net loss is in questions outside the listed types ($-61$ of 447 items). By sub-category, content analysis alone accounts for $-16$ items.
Speaker and voice questions show the same pattern: ahead of Omni-7B on MMAU ($+5$) and 15 items behind on MMAR.
These are the questions where a larger reader helps most (\S\ref{sec:res-scaling}).
\ours{} is also ahead of Omni-7B on MMAR music-theory questions ($+4$ items).
Because every loss is visible in the context, each group points to a specific repair: event-level nodes for counting, finer leaves and scene-level attributes for identity and scenes, and a stronger reader for inference over speech.

\subsection{By benchmark sub-category}
\label{app:subcat}

Tables~\ref{tab:subcat-mmau} and~\ref{tab:subcat-mmar} compare \ours{} (Qwen2.5-7B reader) with Omni-7B on every benchmark sub-category.
The MMAU losses are in acoustic-scene reasoning ($-12$ items), musical texture ($-8$), event-based sound reasoning ($-7$), acoustic source inference, temporal event reasoning and melodic structure ($-6$ each).
The MMAR losses are in environmental perception ($-23$), content analysis ($-16$), counting and statistics ($-15$), correlation analysis ($-12$) and professional knowledge ($-11$).
We are ahead of Omni-7B on MMAU emotion-flip detection ($+5$), music temporal reasoning ($+4$), emotion-state summaries ($+2$) and conversational fact retrieval ($+1$), and level on harmony and chord progressions.

\begin{table}[!htbp]
\centering
\caption{\textbf{MMAU by sub-category}, Qwen2.5-7B reader. Net = items right for us minus items right for Omni-7B.}
\label{tab:subcat-mmau}
\scriptsize
\begin{tabular}{lrrrrr}
\toprule
sub-category & n & ours$\uparrow$ & Omni$\uparrow$ & blind$\uparrow$ & net$\uparrow$ \\
\midrule
Acoustic Scene Reasoning & 48 & 56.2 & 81.2 & 33.3 & -12 \\
Musical Texture Interpretation & 34 & 58.8 & 82.4 & 44.1 & -8 \\
Event-Based Sound Reasoning & 48 & 64.6 & 79.2 & 62.5 & -7 \\
Acoustic Source Inference & 48 & 75.0 & 87.5 & 72.9 & -6 \\
Temporal Event Reasoning & 48 & 58.3 & 70.8 & 25.0 & -6 \\
Melodic Structure Interpretation & 33 & 48.5 & 66.7 & 42.4 & -6 \\
Ambient Sound Interpretation & 47 & 76.6 & 89.4 & 59.6 & -6 \\
Instrumentation & 35 & 71.4 & 85.7 & 48.6 & -5 \\
Emotional Tone Interpretation & 33 & 72.7 & 87.9 & 45.5 & -5 \\
Phonological Sequence Decoding & 48 & 83.3 & 89.6 & 62.5 & -3 \\
Eco-Acoustic Knowledge & 46 & 80.4 & 87.0 & 63.0 & -3 \\
Sound-Based Event Recognition & 46 & 84.8 & 91.3 & 45.7 & -3 \\
Counting & 29 & 48.3 & 55.2 & 41.4 & -2 \\
Musical Genre Reasoning & 34 & 76.5 & 82.4 & 58.8 & -2 \\
Event-Based Knowledge Retrieval & 33 & 72.7 & 75.8 & 60.6 & -1 \\
Phonemic Stress Pattern Analysis & 53 & 54.7 & 56.6 & 34.0 & -1 \\
Socio-cultural Interpretation & 20 & 75.0 & 80.0 & 60.0 & -1 \\
Rhythm and Tempo Understanding & 46 & 54.3 & 56.5 & 56.5 & -1 \\
Dissonant Emotion Interpretation & 35 & 91.4 & 91.4 & 82.9 & +0 \\
Key highlight Extraction & 21 & 90.5 & 90.5 & 61.9 & +0 \\
Multi Speaker Role Mapping & 27 & 100.0 & 100.0 & 14.8 & +0 \\
Lyrical Reasoning & 10 & 90.0 & 90.0 & 60.0 & +0 \\
Harmony and Chord Progressions & 33 & 63.6 & 63.6 & 54.5 & +0 \\
Conversational Fact Retrieval & 22 & 95.5 & 90.9 & 40.9 & +1 \\
Emotion State summarisation & 44 & 59.1 & 54.5 & 31.8 & +2 \\
Temporal Reasoning & 56 & 41.1 & 33.9 & 41.1 & +4 \\
Emotion Flip Detection & 20 & 55.0 & 30.0 & 25.0 & +5 \\
\bottomrule
\end{tabular}
\end{table}

\begin{table}[!htbp]
\centering
\caption{\textbf{MMAR by sub-category}, Qwen2.5-7B reader. Net = items right for us minus items right for Omni-7B.}
\label{tab:subcat-mmar}
\scriptsize
\begin{tabular}{lrrrrr}
\toprule
sub-category & n & ours$\uparrow$ & Omni$\uparrow$ & blind$\uparrow$ & net$\uparrow$ \\
\midrule
Environmental Perception and Reasoning & 149 & 57.7 & 73.2 & 30.9 & -23 \\
Content Analysis & 301 & 62.1 & 67.4 & 44.9 & -16 \\
Counting and Statistics & 99 & 31.3 & 46.5 & 25.3 & -15 \\
Correlation Analysis & 50 & 44.0 & 68.0 & 28.0 & -12 \\
Professional Knowledge and Reasoning & 70 & 47.1 & 62.9 & 34.3 & -11 \\
Speaker Analysis & 48 & 52.1 & 68.8 & 29.2 & -8 \\
Acoustic Quality Analysis & 18 & 22.2 & 61.1 & 22.2 & -7 \\
Culture of Speaker & 50 & 58.0 & 68.0 & 42.0 & -5 \\
Music Theory & 63 & 42.9 & 50.8 & 42.9 & -5 \\
Emotion and Intention & 60 & 55.0 & 61.7 & 30.0 & -4 \\
Aesthetic Evaluation & 8 & 37.5 & 75.0 & 12.5 & -3 \\
Anomaly Detection & 17 & 58.8 & 76.5 & 52.9 & -3 \\
Spatial Analysis & 15 & 40.0 & 53.3 & 46.7 & -2 \\
Temporal Analysis & 28 & 25.0 & 32.1 & 35.7 & -2 \\
Audio Difference Analysis & 8 & 50.0 & 62.5 & 62.5 & -1 \\
Imagination & 10 & 60.0 & 70.0 & 40.0 & -1 \\
\bottomrule
\end{tabular}
\end{table}

\FloatBarrier
\section{Audio LLMs as judges}
\label{app:judge}

\paragraph{Protocol.}
Items are sampled from clips outside the development set, with a fixed seed: 400 from MMAU (stratified by group and sub-category), 400 from MMAR (proportional to group) and 100 multi-hop items per SAKURA track.
The judge hears the first 60\,s of the clip and reads the served nodes as the reader receives them, without the transcript and without the question.
It answers with one digit, and the score is the expected value of its probabilities over the digits 1--5, so no output needs parsing.
The swap control uses the nodes of another sampled clip of the same benchmark from a different category (disjoint modalities, e.g.\ music for speech; Animal for the SAKURA speech tracks), chosen to share as few nodes as possible with the real ones (mean node overlap below 1\%).
Confidence intervals are 95\% bootstrap intervals over items (2{,}000 resamples).

\paragraph{Prompt.}
The system message and the user message are given below; \texttt{\{nodes\}} is the node list of the item (or of the swapped clip), and the audio is attached to the user message.

\begin{tcolorbox}[colback=gray!4, colframe=gray!60, boxrule=0.5pt, arc=2pt, left=4pt, right=4pt, top=2pt, bottom=2pt, fontupper=\scriptsize]
\textbf{System:} You are an expert audio analyst auditing the output of an automatic audio-analysis system. Listen to the whole recording before answering, and base every judgement on what you hear.\\[3pt]
\textbf{User:} An automatic system analysed this recording and matched it to nodes of an audio knowledge graph, written as ``attribute: value'' pairs; a value in parentheses is a more specific label. The system can be wrong.\\[2pt]
Nodes:\\
\{nodes\}\\[2pt]
Are these nodes related to what is actually audible in the recording?\\
5 = every node matches something audible in the recording\\
4 = nearly all nodes match; one is doubtful or too vague\\
3 = the main content is captured, but several nodes do not match\\
2 = most nodes do not match; the main source, speaker or music is wrong\\
1 = the nodes describe a different recording\\[2pt]
Answer with a single digit from 1 to 5 and nothing else.
\end{tcolorbox}

All three judges rank the served nodes above the swapped ones on most items (Figure~\ref{fig:judge}).
They differ in scale: Audio Flamingo 3 rates even the swapped nodes above 3, and Kimi-Audio rates the served nodes lowest, most of all on the SAKURA speech tracks (language and gender).
Averaging the three judges after standardising each one's scores gives the same ordering on every benchmark.

\FloatBarrier
\section{Example contexts}
\label{app:examples}

Figure~\ref{fig:examples} shows three MMAU items with the complete context given to the reader.
More examples, with the router's path through the tree chunk by chunk, are on the project page.

\begin{figure}[!htbp]
\centering
\begin{tcolorbox}[colback=teal!4, colframe=teal, boxrule=0.6pt, arc=2pt, left=4pt, right=4pt, top=2pt, bottom=2pt, fontupper=\scriptsize\sloppy, title={\small Correct: \textit{What mechanical issue is likely occurring with the vehicle?}}, fonttitle=\small]
Options: (a) Tire puncture \; (b) Engine misfire \; \textbf{(c) Engine knocking} \; (d) Brake failure \hfill Prediction: \textbf{(c)}\\[2pt]
\texttt{sound scene: construction site; sound machine: engine (Engine), \textbf{engine (Engine knocking)}, mechanisms (Mechanisms); sound action material: impact (keys hitting), impact; sound proximity: close to the microphone.}
\end{tcolorbox}
\vspace{-2pt}
\begin{tcolorbox}[colback=teal!4, colframe=teal, boxrule=0.6pt, arc=2pt, left=4pt, right=4pt, top=2pt, bottom=2pt, fontupper=\scriptsize\sloppy, title={\small Correct: \textit{Given the sounds, what type of weather event can be inferred?}}, fonttitle=\small]
Options: \textbf{(a) Thunderstorm} \; (b) Clear skies \; (c) Heatwave \; (d) Snowstorm \hfill Prediction: \textbf{(a)}\\[2pt]
\texttt{sound scene: park; sound nature: water (Rain), water (Water), \textbf{thunderstorm (Thunder)}, thunderstorm (Thunderstorm); sound action material: liquid flow (water rushing), air flow; sound proximity: close to the microphone.}
\end{tcolorbox}
\vspace{-2pt}
\begin{tcolorbox}[colback=burnt!5, colframe=burnt, boxrule=0.6pt, arc=2pt, left=4pt, right=4pt, top=2pt, bottom=2pt, fontupper=\scriptsize\sloppy, title={\small Wrong, and traceable: \textit{Identify the source of the waterfall sound.}}, fonttitle=\small]
Options: \textbf{(a) Waterfall} \; (b) Rain \; (c) Ocean waves \; (d) River \hfill Prediction: (b)\\[2pt]
\texttt{sound scene: park; sound impact: liquid (Liquid), liquid (Boiling); sound household: frying food (Frying (food)), toilet flush; sound machine: mechanisms (Tick); sound nature: \textbf{water (Rain)}, water (Water); sound action material: liquid flow (water rushing).}\\[2pt]
\textrm{The router named \texttt{Rain} and no waterfall node; the reader followed it. The error is located in one node of the \texttt{water} class, which is where a fix (a \texttt{Waterfall} leaf with training clips) belongs.}
\end{tcolorbox}
\caption{\textbf{Interpretable by design.} Real MMAU items with the complete tree context given to the frozen Qwen2.5-7B (no transcript is given for these clips: no chunk heard speech or singing). Bold marks the node that decides the answer.}
\label{fig:examples}
\end{figure}

\FloatBarrier
\section{Cost comparison}
\label{app:cost}

Table~\ref{tab:cost} gives the cost of \ours{} and the audio LLMs discussed in \S\ref{sec:res-cost}.
Qwen3-Omni needs the fewest MACs despite its 35.3B parameters, because it is a mixture-of-experts model that runs about 3B parameters per token, but its per-item time is still the highest (0.93\,s).

\begin{table}[!htbp]
\centering
\caption{\textbf{Cost} of \ours{} (Qwen2.5-7B reader) and audio LLMs. MACs and seconds are per MMAU item on one A100 GPU; MACs count executed operations, so a mixture-of-experts model counts only its active parameters.}
\label{tab:cost}
\small
\setlength{\tabcolsep}{3.5pt}
\resizebox{\linewidth}{!}{%
\begin{tabular}{lccccc}
\toprule
& total params$\downarrow$ & trained params$\downarrow$ & audio training data$\downarrow$ & MACs / item$\downarrow$ & s / item$\downarrow$ \\
\midrule
\textbf{\oursacc{} (ours)} & 10.50B & 7.7M & $\approx$0.2M clips & 5.3T & 0.57 \\
Qwen2.5-Omni-7B & 10.73B & 10.73B & 300B tokens & 2.7T & 0.12 \\
Audio Flamingo 3 & 8.27B & 8.27B & 9.6M QA + 13.2M captions & 3.3T & 0.11 \\
Qwen3-Omni-30B-A3B & 35.3B (3B active) & 35.3B & 20M h + 0.77T tokens & 0.76T & 0.93 \\
\bottomrule
\end{tabular}}
\end{table}

\FloatBarrier